\documentclass[aps,prx,twocolumn,superscriptaddress,longbibliography]{revtex4-1}

\usepackage[english]{babel} 
\usepackage[latin1]{inputenc}
\usepackage[usenames,dvipsnames]{color}
\usepackage{graphicx}
\usepackage{amsmath,amssymb,amsfonts}
\usepackage[colorlinks=true,linkcolor=blue,urlcolor=blue,citecolor=blue]{hyperref}
\usepackage[percent]{overpic}

\begin{document}

\title{Ultrafast charging in a two-photon Dicke quantum battery}

\author{Alba Crescente}
\affiliation{Dipartimento di Fisica, Universit\`a di Genova, Via Dodecaneso 33, 16146, Genova, Italy}
\affiliation{CNR-SPIN, Via Dodecaneso 33, 16146, Genova, Italy}
\author{Matteo Carrega}
\affiliation{CNR-SPIN, Via Dodecaneso 33, 16146, Genova, Italy}

\author{Maura Sassetti}
\affiliation{Dipartimento di Fisica, Universit\`a di Genova, Via Dodecaneso 33, 16146, Genova, Italy}
\affiliation{CNR-SPIN, Via Dodecaneso 33, 16146, Genova, Italy}
\author{Dario Ferraro}
\affiliation{Dipartimento di Fisica, Universit\`a di Genova, Via Dodecaneso 33, 16146, Genova, Italy}
\affiliation{CNR-SPIN, Via Dodecaneso 33, 16146, Genova, Italy}

\begin{abstract}
We consider a collection of two level systems, such as qubits, embedded into a microwave cavity as a promising candidate for the realization of high power quantum batteries. In this perspective, the possibility to design devices where the conventional single-photon coupling is suppressed and the dominant interaction is mediated by two-photon processes is investigated, opening the way to an even further enhancement of the charging performance. 
By solving a Dicke model with both single- and two-photon coupling we determine the range of parameters where the latter unconventional interaction dominates the dynamics of the system leading to better performances both in the charging times and average charging power of the QB compared to the single-photon case.
In addition, the scaling of the maximum stored energy, fluctuations and charging power with the finite number of qubits $N$ is inspected.
While the energy and fluctuations scale linearly with $N$, the quadratic growth of the average power leads to a relevant improvement of the charging performance of quantum batteries based on this scheme with respect to the purely single-photon coupling case.
 Moreover, it is shown that the charging process is progressively faster by increasing the coupling from the weak to the ultra-strong regime.
\end{abstract}

\maketitle

\section{Introduction}

Recent advances in quantum technologies~\cite{Riedel17, Zhang19_qt, Raymer19, Sussman19} allow to access and manipulate microscopic systems at the single-atom or single-photon level with very high precision. Here, quantum features, such as phase-coherence or entanglement, play an essential role both from a fundamental and an applicative point of view. Great interest revolves around the possibility to exploit quantum resources to bypass bottlenecks posed by classical physics, improving performances both in computation and information schemes~\cite{DiVincenzo95}.

These concepts have been recently addressed also in the fast developing field of quantum thermodynamics~\cite{Esposito09, Vinjanampathy16, Bera19, DePasquale18, Carrega19, Benenti17, Pekola15, Levy12, Campisi16}, where energy flows, and related work, between microscopic systems are considered in a fully quantum setting. In this context, so-called ``Quantum Batteries'' (QBs) have been proposed~\cite{Alicki13, Hovhannisyan13, Binder15, Campaioli17, Campaioli18, Bhattacharjee20}, and are currently under experimental investigations, as small devices where energy can be stored in faster or more efficient ways compared to their classical counterpart. Several studies focussed on the understanding of how the presence of quantum coherences or entanglement can affect energy storage in individual~\cite{Andolina18, Crescente20, Carrega20, Mohan20, Bai20, TRYang20} and many-body QBs~\cite{Binder15, Le18, Farre20, Rossini19, Rosa19, Zhang19, Huang20, Ghosh20, Tabesh20, Andolina19b}, with the aim to find and demonstrate the so-called quantum advantage for QBs. Experimentally feasible nanodevices based on engineered two-level systems (TLSs) realized by means of superconducting qubits~\cite{Devoret13} or semiconducting quantum dots~\cite{Singha11} have been also put forward.

Circuit-QED~\cite{Devoret13, Schoelkopf08} represents another interesting platform where light-matter interactions can be engineered and exploited to build QBs playing with the interplay between solid-state devices and photonic degrees of freedom. Here, a paradigmatic model is the so-called Dicke model~\cite{Dicke54}, where a collection of TLSs interacts with a single-photon mode of a cavity, representing one of the most studied model in quantum optics. Despite its simplicity, it is still of great interest, with the recent realization of coupled systems even at ultrastrong coupling (USC)~\cite{Niemczyk10, Bayer17, Kockum19, FDiaz19, Macha14, Kakuyanagi16, Yoshihara17, Stufler06}. Further, the possibility to host super-radiant phase transition~\cite{Hepp73, Wang73, Emary03} is an on going subject of research. Recently, QBs based on the Dicke model have been introduced~\cite{Ferraro18}, assuming a conventional dipole coupling between the TLSs and photon radiation of the cavity, where it has been reported that a $\sqrt{N}$ speed up in the charging process can be achieved, compared to an analog parallel charging scheme, where each TLS is coupled with its own cavity. This enhancement of the performance is ultimately related to the renormalization of the coupling constant due to many-body interaction~\cite{Fink09}. Moreover, the possibility to increase the electrostatic capacity of the system exploiting the interplay between TLS-TLS interaction and coupling with the radiation has been also investigated~\cite{Ferraro19}.

Some recent proposals, based on trapped ions~\cite{Felicetti15} or superconducting flux qubits~\cite{Felicetti18} have suggested the possibility to suppress the dipole contribution, linear in the photon coupling, and to enhance the two-photon coupling. This configuration is predicted to lead to new and interesting physics, enhancing the visibility of a super-radiant phase transition and hinting at a spectral collapse at high coupling~\cite{Emary02, Dolya09, Chen12, Peng17, Garbe17, Garbe19}.

It is therefore natural to ask if this two-photon configuration can lead to further improvement of the functionality of a QB. In the present paper we address this issue comparing a single- and two-photon coupling. By numerically solving a Dicke Hamiltonian with both contributions we determine in which regime of the parameters the single-photon coupling has only marginal effect on the physics and the two-photon contribution dominates. This analysis can have relevant experimental implications, leading to less strict constraints for the observation of purely two-photon effects. We investigate the behavior of physically relevant quantities in characterizing a QB such as stored energy, averaged charging power and energy fluctuations. This latter figure of merit is frequently neglected even if it can strongly affect the functionality of the considered devices~\cite{Friis18, Pintos20}. 
The main result of this study is that a two-photon interaction leads to better performances in terms of charging times and charging power of the QB compared to the single-photon case.
Moreover, a two-photon coupling can lead to a charging $N$ times faster with respect to the parallel charging case and an enhancement of the averaged charging power which scales as $N^{2}$, with an evident advantage with respect to the single-photon case. This improvement is achieved both in the weak coupling regime, where the main physics of the model can be investigated relying on the rotating-wave approximation~\cite{Felicetti15, Felicetti18}, and in the USC regime, where the coupling with the radiation is a consistent fraction of the TLS energy gap.

The paper is organized as follows. In Section~\ref{sec:model} we introduce a generalized Dicke model that takes into account both single- and two-photons coupling. We also define the relevant physical quantities (stored energy, averaged charging power and energy fluctuation) needed in order to properly characterize the behavior of the QBs. Section~\ref{sec:charging} is devoted to the analysis of the two-photon contribution in comparison with the single-photon one in order to identify the parameter regimes where an advantage in term of charging time and averaged charging power can be observed. The scaling of the various quantities as a function of the number of TLSs is reported in Section~\ref{sec:power}. Section~\ref{sec:conclusions} is devoted to the conclusions. Additional details on the exact diagonalization procedure are given in Appendix~\ref{AppA}, a comparison between our numerical results in the weak coupling regime and the ones obtained considering the simplified Tavis-Cummings (TC) model~\cite{TC1, TC2}, rotating wave approximation of the Dicke model, are presented in Appendix~\ref{AppB}, while a discussion of the parallel charging in the weak coupling regime is provided in Appendix~\ref{AppC}.

%%%%%%%%%%%%%%%%%%%%%%%%%%%%%%%

\section{Model \label{sec:model}}

\begin{figure}[h!]
\centering
\includegraphics[width=0.44 \textwidth]{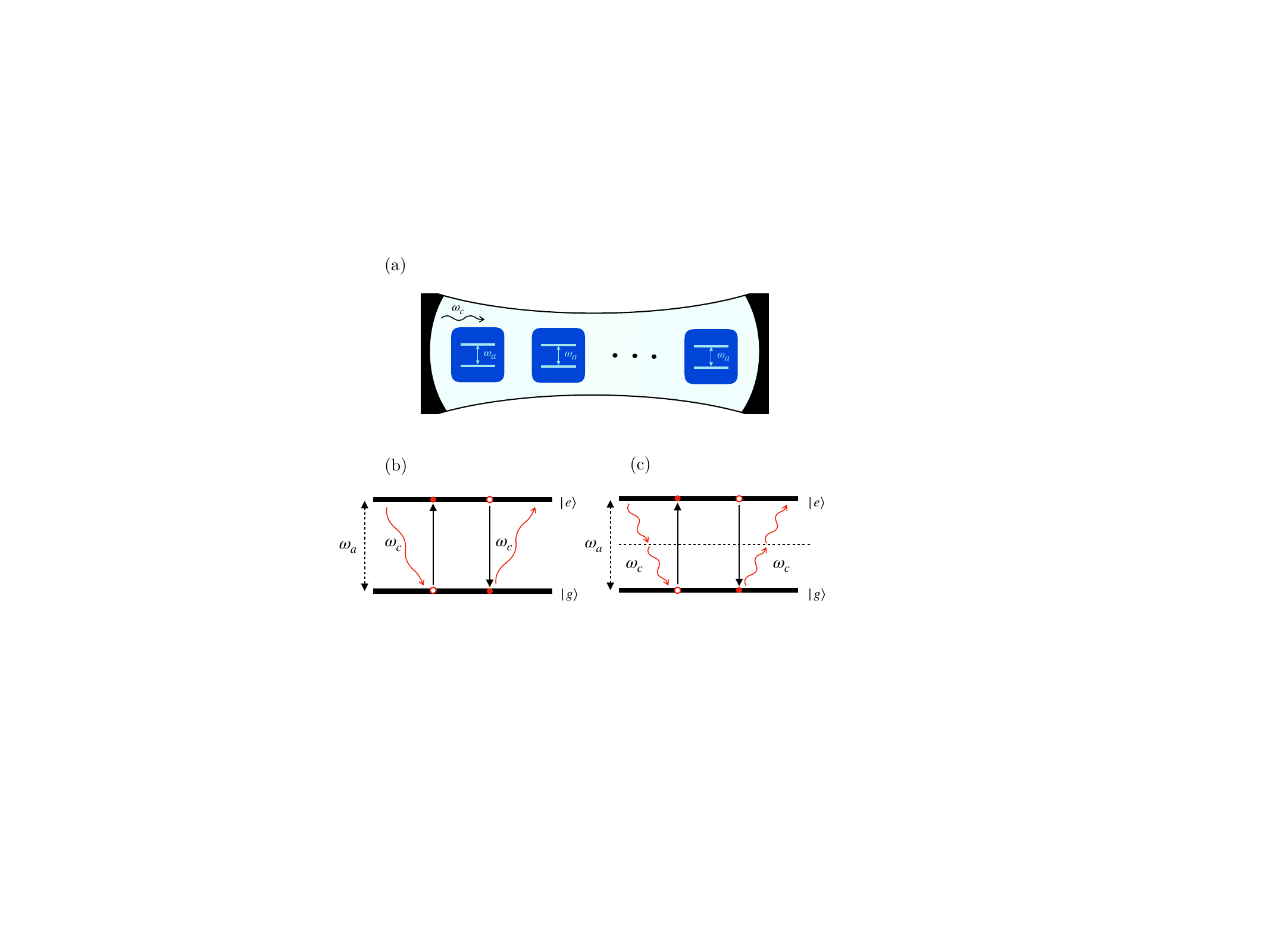}
\caption{Scheme of a QB where a set of $N$ identical and independent TLSs with energy separation $\omega_a$ interact with a unique cavity mode of frequency $\omega_c$ via photon coupling (a). Processes of creation and destruction of photons involved in the single-photon (b) and in the two-photon Dicke regimes (c) in the resonant conditions $\omega_{a}=\omega_{c}$ and $\omega_{a}=2 \omega_{c}$ respectively.}
\label{Figure1}
\end{figure}

We consider a QB modeled as a set of a finite number $N$ of identical and independent TLSs coupled to a unique cavity mode (see Fig.~\ref{Figure1} (a)).

The system can be described using the so-called Dicke model~\cite{Dicke54}. In particular, we want to consider $N$ TLSs coupled to a single cavity mode via both a single-photon~\cite{Ferraro18, Andolina18} and a two-photon coupling~\cite{Garbe17, Garbe19}, for which the Hamiltonian is (hereafter we set $\hbar = 1$)

\begin{eqnarray} 
\label{D2ph} H&=& \omega_ca^\dagger a + \omega_a J_z+\theta(t)H_I^{1ph} + \theta(t)H_I^{2ph},
\end{eqnarray}

\noindent where

\begin{eqnarray}
\label{H_1ph}
H_I^{1ph}&=&g_1J_x(a^\dagger+a)  \\
\label{H_2ph}
H_I^{2ph}&=&g_2J_x[(a^\dagger)^2+(a)^2]
\end{eqnarray}

\noindent represent the interaction terms in the case of a single- and two-photon coupling, respectively.

Here, $\omega_c$ is the frequency of the photons in the cavity, $\omega_a$ is the energy splitting between the ground state $|g\rangle$ and excited state $|e\rangle$ of each TLS, $g_1$ and $g_2$ are the coupling strengths for the single-photon and the two-photon interactions respectively. 
The notation 
\begin{equation} 
J_\alpha= \frac{1}{2} \sum_{i=1}^N \sigma_i^\alpha 
\end{equation}
with $\alpha=x,y,z$ indicates the components of a pseudo-spin operator expressed in terms of the Pauli matrices of the $i$-th TLS. Finally, $a$ ($a^\dagger$) annihilates (creates) a photon in the cavity, as represented in Fig.~\ref{Figure1} (b) and $a^2$ ($(a^\dagger)^2$) annihilates (creates) a pair of photons in the cavity, as in Fig.~\ref{Figure1} (c).
We also assume that the coupling between the TLSs and the cavity, used as a charger for the QB ($J_{z}$), is switched on at $t=0$, as indicated by the $\theta(t)$ functions in Eq.~(\ref{D2ph}).

Moreover, we notice that the interaction term in Eq. (\ref{H_1ph}) is related to a conventional linear coupling with the cavity electric field~\cite{Schleich_book}, while the one in Eq. (\ref{H_2ph}) is quadratic. In experiments realized in the framework of circuit quantum electrodynamics,
the dominant light-matter interaction contribution is usually the dipolar one which is linear in the photon annihilation/creation operators~\cite{Chen07, Nataf10,Nataf10b}. However theoretical proposals, with the aim of enhancing the two-photon contribution (quadratic in the electric field), have been recently presented. It has been shown that by properly engineering the device configuration it is possible to suppress, or even eliminate, one-photon contribution, thus promoting the two-photon as the dominant one~\cite{Felicetti15,Felicetti18,Kockum19}.
In particular, Refs.~\cite{Felicetti15, Felicetti18} have discussed the case of trapped ions subject to a bichromatic driving and the case of a flux qubit coupled with a symmetric dc SQUID respectively. In both these scenarios, the authors claim the possibility to span from the weak coupling to the USC regime in realistic experimental configurations. 
It is important to underline that the technologies introduced in these works are commonly used for the implementation of qubits and have reached great experimental control opening to the possibility of actually realizing devices showing a relevant two-photon coupling in the near future.

Notice that in Eq.~(\ref{D2ph}) we have used the notation already considered in Refs.~\cite{Ferraro18,Andolina19}. However, other definitions for the coupling constants (e.g. rescaled with respect to the number of TLSs $N$) can be found in literature~\cite{Garbe17, Garbe19, Felicetti15, Fink09}. While the former are often used in context dealing with finite size systems with small number of TLSs, the latter are more conventional for studying the large $N$ limit, being consistent with the expected thermodynamic limit $N\to\infty$~\cite{Farre20}.

Here, we consider the dynamics of a closed quantum system. Interactions with the external environment can lead to relaxation and loss of photons in the cavity characterized by typical time scales $t_{r}$ and $t_{\gamma}$ respectively~\cite{Haroche_Book, Carrega20, Devoret13, Paladino08, Wendin17} therefore, in the following, we restrict the analysis to evolution times such that $t\ll t_{r}, t_{\gamma}$ where dissipation effects can be neglected. According to the acquired experimental level of control of the qubits this condition is typically fulfilled in state of the art circuit quantum electrodynamics devices~\cite{Devoret13}. It is important to underline that usually $t_r > t_\gamma$, so experimentally it is important to consider stable cavities. Depending on the considered technology typical values of $t_\gamma$ range from $\sim$ $ \mu$s in transmon qubits~\cite{Devoret13, Koch07} to $\sim$ ms in trapped ions~\cite{Talukdar16}, which need to be compared to the effective Rabi frequency characterizing the time evolution of the system.

In addition, we focus on the two resonant regimes, namely $\omega_a=\omega_c$ (see Fig.~\ref{Figure1} (b)) where one expects a dominant contribution from the single-photon process and $\omega_a=2\omega_c$ (see Fig.~\ref{Figure1} (c)) where the two-photon interaction should be more relevant. Off-resonance cases ($\omega_a\neq\omega_c$, $\omega_a\neq2\omega_c$) will not be discussed due to the fact that they are characterized by a less efficient energy transfer between the cavity and the TLSs~\cite{Schleich_book}.
 
This allows us to consider initial states of the form 

\begin{equation}
\label{psi2ph} |\psi(0)\rangle = |s N\rangle \otimes \underbrace{|g,...,g\rangle}_{N}.
\end{equation}
Here, $s=1$ for the first resonant case and $s=2$ for the second, the $N$ TLSs are prepared in the ground state $| g\rangle$ and the cavity mode is in the $s N$ Fock states. It is worth to stress that, even if other initial states can be studied in analogy with what done in Ref.~\cite{Crescente20}, this particular choice guarantees to have initial states carrying the minimum number of photons necessary for the radiation to work as a charger for the QB in the single-photon or in the two-photon resonance.

\subsection{Figures of merit}

To characterize a QB we study the total energy that can be stored, the corresponding charging time and the average charging power, namely the energy stored in a given time interval. Moreover, we also consider energy fluctuations, to see how their detrimental effects can influence the functionality of the QB~\cite{Friis18, Crescente20}.

\subsubsection{Stored energy and charging power}
At time $t$ the energy stored is given by~\cite{Ferraro18}

\begin{eqnarray}\label{E(t)} E(t)&=& \omega_a [\langle \psi(t) | J_z | \psi(t) \rangle
- \langle \psi(0) | J_z | \psi(0) \rangle],  \end{eqnarray}

\noindent where $|\psi(t)\rangle=e^{-i H t}|\psi(0)\rangle$. 

We also define the average charging power at time $t$ as~\cite{Andolina18}

\begin{equation}
\label{P(t)} P(t)= \frac{E(t)}{t}. 
\end{equation}

In both cases we are looking for the fastest possible storage of energy and the greater charging power (occurring at times $t_E$ and $t_P$ respectively). We then define~\cite{Ferraro18, Andolina18}

\begin{eqnarray}
\label{Emax} E_{max} &\equiv& \underset{t}{ \text{max}}[E(t)]\equiv E(t_E)\\
\label{Pmax} P_{max} &\equiv& \underset{t}{\text{max}}[P(t)]\equiv P(t_P).
\end{eqnarray}

We underline that the above equations strongly depend on the value of the coupling strengths $g_1$, $g_2$ as it will be clearer in the following.

\subsubsection{Energy fluctuations}
As another useful quantifier of the QB performance we consider the quantum fluctuations of the energy. To do so we  analyze the fluctuations between the initial and final time of the charging process represented by the correlator~\cite{Friis18}

\begin{eqnarray}
\label{Sigma} 
\Sigma^2(t) &=& \omega_a^2\bigg[\sqrt{\langle J_z^2(t)\rangle-(\langle J_z(t)\rangle)^2} \nonumber \\
&-&\sqrt{\langle J_z^2(0)\rangle-(\langle J_z(0)\rangle)^2}\bigg]^2,
\end{eqnarray}

\noindent where $J_z(t)$ is the Heisenberg time evolution of the operator $J_z$ and the averages are taken with respect to the initial state in Eq. (\ref{psi2ph}).
We emphasize that $\Sigma(t)$ is related to the inverse of the so called reverse quantum speed limit which can be used to characterize the discharging of the QB~\cite{Mohan20}. 
Moreover another correlator~\cite{Friis18, Crescente20}, representing fluctuations at equal time, can be considered, but for the chosen initial states it is identical to $\Sigma(t)$, as also explained in Ref.~\cite{Crescente20}.

\subsection{Numerical approach}

As already discussed in the single-photon case~\cite{Ferraro18}, in order to evaluate the energy, power and fluctuations in Eqs.~(\ref{E(t)}), (\ref{P(t)}) and (\ref{Sigma}) starting from the Dicke model in Eq.~(\ref{D2ph}) we need to use a numerical approach (see Appendix \ref{AppA} for more details). The reason for this is that the Dicke model doesn't conserve the number of excitations, as it can be clearly seen from the interaction term $g_1\theta(t)J_x(a^\dagger+a)+g_2\theta(t)J_x[(a^\dagger)^2+(a)^2]$ of Eq.~(\ref{D2ph}) which contains counter-rotating terms of the form $a^\dagger J_+$, $(a^\dagger)^2J_+$ and $aJ_-$, $a^2J_-$. However, we notice that $J^2= J_x^2+J_y^2+J_z^2$ is a conserved quantity for the Hamiltonian in Eq.~(\ref{D2ph}). This allows us to work in the basis $|n,j,m\rangle$, where $n$ is the number of photons, $j(j+1)$ is the eigenvalue of $J^2$ and $m$ is the eigenvalue of $J_z$. Within this notation the initial state in Eq.~(\ref{psi2ph}) can be written as
\begin{equation} 
|\psi(0)\rangle=|s N, N/2, -N/2\rangle. 
\end{equation}

\begin{figure*}[ht]
\centering
\includegraphics[width=1.8\columnwidth]{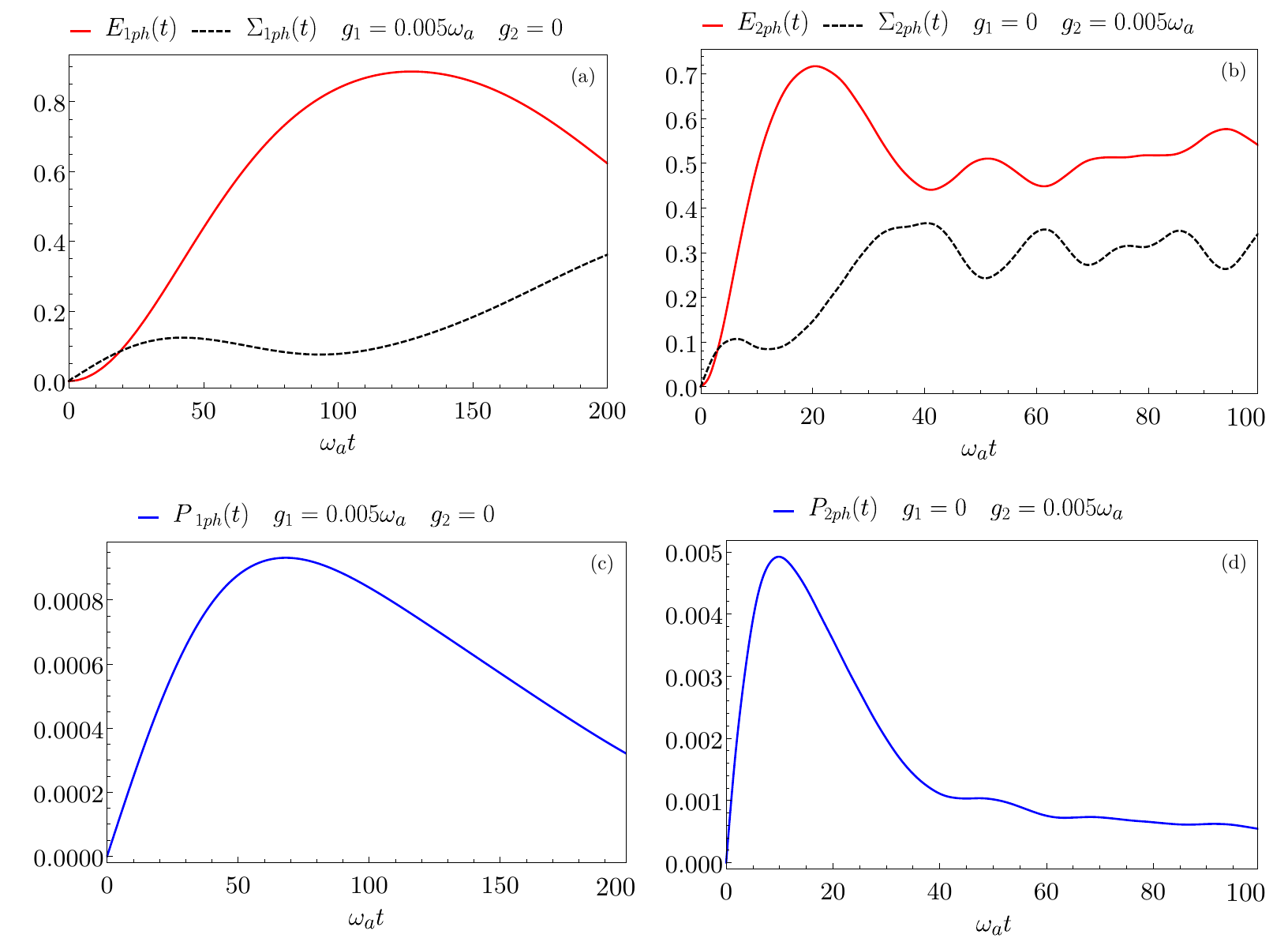}
\caption{Behavior of the stored energy and its fluctuations (in units of $N\omega_a$) as a function of the time (in units of $\omega^{-1}_{a}$) for the single-photon Dicke regime at $g_1=0.005\omega_{a}$ and $g_2=0$ (a) and for the two-photon Dicke regime at $g_1=0$ and $g_2=0.005\omega_{a}$ (b). Behavior of $P(t)$ (in units of $N^2\omega_a^2$) as a function of the time (in units of $\omega^{-1}_{a}$) for the single-photon Dicke regime ($g_1=0.005\omega_{a}$ and $g_2=0$) (c) and for the two-photon Dicke regime ($g_1=0$ and $g_2=0.005\omega_{a}$) (d). All the plots show the case $N=10$. Different timescales in the panels have been used in order to better determine the position of the maxima of the various quantities in the two cases.}
\label{Figure2}
\end{figure*}

Even in this basis the numerical problem we have to solve is rather difficult, and it requires, at least in principle, an infinite Hilbert space, because the Dicke Hamiltonian is not bounded from above. Within our finite size numerical diagonalization, we have estensivly checked the numerical convergence of the results (energy, averaged charging power and energy fluctuations) and verified that, even in the worst case scenario of large $N$, setting the maximum number of photons to $N_{ph}$=$4N$ there is a difference below $10^{-5}$ between the results at $N_{ph}+1$ and the one at $N_{ph}$~\cite{Ferraro18, BM11}. This allows us to chose for all the following plots $N_{ph}=4N$.

%%%%%%%%%%%%%%%%%%%%%%%%%%%%%%%%%%%%%%%%%%%%

\section{Improved charging via two-photon coupling \label{sec:charging}}

Here we describe our main results obtained using the fully numerical approach just introduced. We present an analysis of different regimes for the Dicke model, starting from weak coupling and then considering the interesting USC regime, where we expect a faster charging and a further enhancement of the average charging power in analogy with what observed for a purely single-photon coupling~\cite{Ferraro18}. 
For sake of clarity, we will compare the two-photon case, $g_1=0$ and $\omega_a=2\omega_c$ in Eq.~(\ref{D2ph}), with the charging performance of the single-photon one, $g_2=0$ and $\omega_a=\omega_c$ in Eq.~(\ref{D2ph}), already discussed in Ref.~\cite{Ferraro18}. In each Subsections we will also consider how the presence of the $g_1$ interaction can affect the $g_2$ contribution, taking into account the total Hamiltonian in Eq.~(\ref{D2ph}) in the two-photon resonance regime ($\omega_a=2\omega_c$).

\subsection{Weak coupling regime \label{sec:WC}}

We first analyze what happens for the two different cases considering small couplings $g_i \ll \omega_{a}$, with $i=1,2$ denoting the  single- and two-photon coupling constant respectively. 
In the following analysis we consider  $g_i=0.005\omega_{a}$ as a representative value, however other coupling constants in this regime lead to similar features.
In this limit we can compare our results with the one obtained within the TC model~\cite{TC1, TC2}, rotating wave approximation of the Dicke model. More details are reported in Appendix~\ref{AppB}.

In Fig.~\ref{Figure2} (a) and (b) we report the energy $E(t)$ and its fluctuations $\Sigma(t)$ at a given number of TLSs $N=10$ for a purely single- and two-photon coupling, respectively. Notice that, throughout the paper, we will consider the (constant) energy scale $\omega_a$ as a reference of all energy (and time) scales.
From the plots we can observe that the maximum of the stored energy, defined in Eq.~(\ref{Emax}), of the single-photon Dicke case ($E_{max}^{(1ph)}/N\omega_a\sim 0.886$) is higher than the two-photon one ($E_{max}^{(2ph)}/N\omega_a\sim 0.710$), however the charging time in the first case is $\omega_at_{E}^{(1ph)}=127.33$, while in the latter is way faster ($\omega_at_{E}^{(2ph)}=19.17$). Therefore, introducing the ratio

\begin{equation} 
\mathcal{R}=\frac{t_E^{(1ph)}}{t_E^{(2ph)}}
\label{ratio_R}
\end{equation}

\noindent one achieves here the value $\mathcal{R}\sim 6.64$. 
We want to underline that this advantage in the charging times of the QB is achieved both for small and large $N$. In the former case this is ultimately related to the bosonic nature of the photon interaction (cavity), see also Appendix~\ref{AppC} for the illustrative case $N=1$, while in the latter also the collective behaviour of the $N$ TLSs coupled to the single cavity plays a role.
The latter point will be better discussed in Section~\ref{sec:power}.

Energy quantum fluctuations are reported in Fig.~\ref{Figure2} (a) and (b) (dashed curves). For the functionality of the QBs it is important to consider the value of $\Sigma(t)$ at the time $t_E$ where the maximum of the energy occurs, defined as
\begin{equation}
\label{barS} \bar{\Sigma} \equiv \Sigma(t_E). 
\end{equation}
We can observe that in the single-photon case $\bar{\Sigma}^{(1ph)}/N\omega_{a}=0.119$, while in the two-photon one $\bar{\Sigma}^{(2ph)}/N\omega_{a}=0.141$. This result is directly linked to the fact that, due to the interaction between the $N$ TLSs and the cavity mode, the two cases do not reach the full charging of the QB. Indeed, according to what reported in Ref.~\cite{Crescente20}, energy fluctuations are absent only if the total charge of the QB is reached. 

From Fig.~\ref{Figure2} we can see an enhancement of the maximum of the averaged charging power (defined in Eq.~(\ref{Pmax})) in the two-photon regime $P_{max}^{(2ph)}/ N^2\omega_a^2\sim 0.005$ (panel (d)) compared to the one in the single-photon case where $P_{max}^{(1ph)}/ N^2\omega_a^2\sim0.0008$ (panel (c)) with a ratio
\begin{equation} 
\mathcal{L}=\frac{P_{max}^{(2ph)}}{P_{max}^{(1ph)}}
\label{ratio_L}
\end{equation}
given here by $\mathcal{L}\sim 6.25$, proving the better performances of the two-photon Dicke regime~\cite{Note_power}.

We conclude this section analyzing how the two-photon Dicke regime at resonance can be affected by the single-photon contribution. We then consider the complete Hamiltonian in Eq.~(\ref{D2ph}) and we define the ratio
\begin{equation}
\label{k} 
k=\frac{g_1}{g_2}. 
\end{equation}

\begin{figure}[h!]
\centering
\includegraphics[width=0.9\columnwidth]{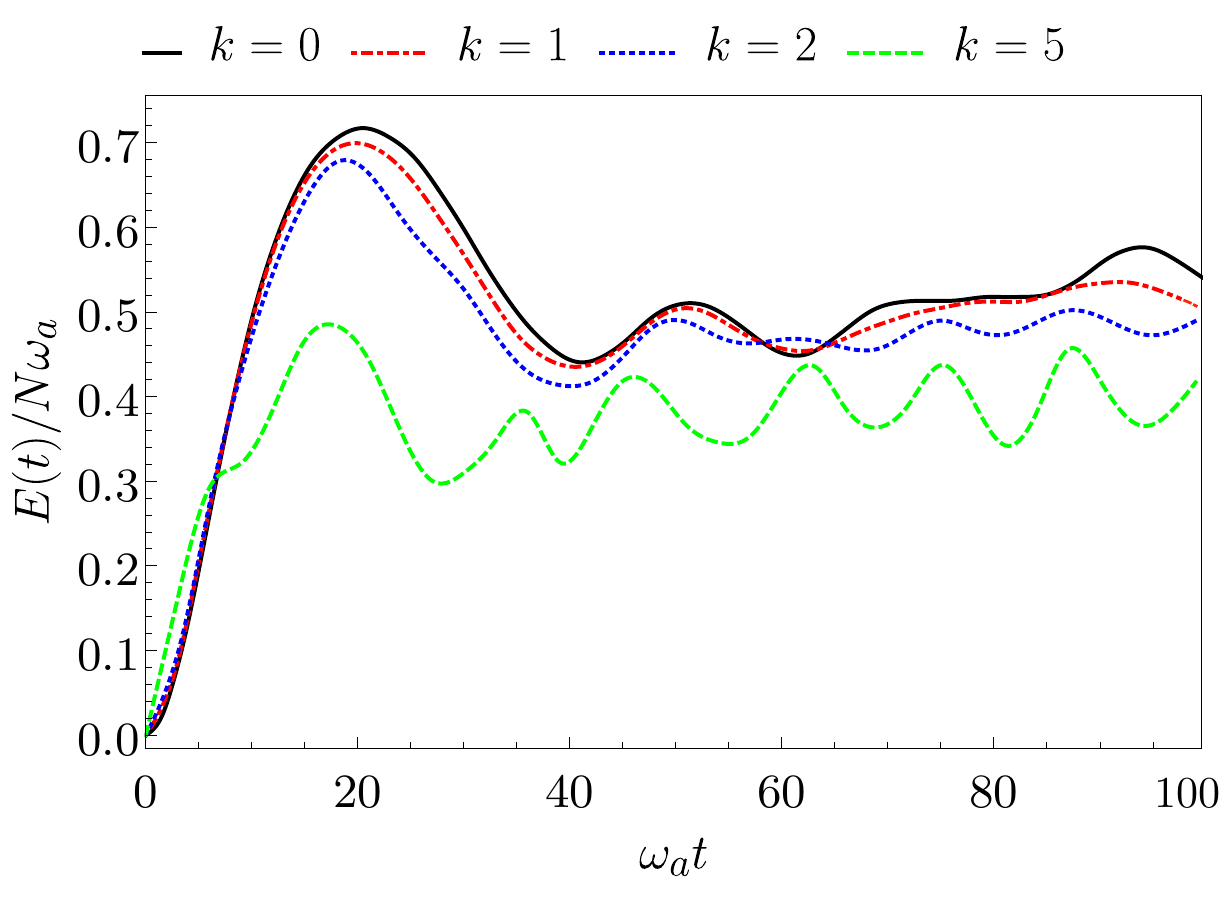}
\caption{Behavior of $E(t)$ (in units of $N\omega_a$) as a function of the time (in units of $\omega^{-1}_{a}$) for the Dicke model in Eq.~(\ref{D2ph}) at the two-photon resonance ($\omega_a=2\omega_c$) for different values of $k$. Other parameters are $g_2=0.005\omega_{a}$ and $N=10$.}
\label{Figure3}
\end{figure}

In Fig.~\ref{Figure3} we report the behavior of the $E(t)$ for different values of $k$ at fixed $g_2=0.005\omega_{a}$. 

We start from the case already considered in Fig.~\ref{Figure2} (b) ($g_1=0$), where the single-photon coupling is absent (black curve) and we examine what happens when we increase the strength of the single-photon interaction. From the plots we can see that when the two couplings are identical (dash-dotted red curve at  $k=1$) we obtain roughly the same behavior of a purely two-photon contribution. Increasing $g_1$ (dotted blue curve at $k=2$) differences in the maximum of the energy start to emerge. At even higher values of $g_{1}$ (dashed green curve at $k=5$) the discrepancy with respect to the two-photon case in Fig.~\ref{Figure2} (a) is marked and the maximal stored energy is lower. This analysis allows us to understand that, in the weak coupling regime ($g_i\ll \omega_a$), for ratios up to $k\approx2$ ($g_1\approx 2g_2$), the system behaves mainly as a two-photon Dicke case (with associated better performances). This represents an interesting fact for future experimental implementation, meaning that it is not required to completely "switch off" the single-photon contribution for the system to effectively work in the resonant two-photon regime.

Notice that for the value $N=10$ considered in this Section a universal behavior of the discussed physical quantities is reached (see below). However, one needs to keep in mind that preserving the coherence over such a large number of cavity photons and TLSs is a challenging task from the experimental point of view~\cite{Hofheinz08, Fink09}.
This is because the lifetime of a both the Fock state $|N\rangle$ and the many-body state of the TLSs decrease with the dimension of the system~\cite{Lu89}.

\subsection{Ultra-strong coupling regime \label{sec:USC}}

We now discuss coupling strengths in the USC regime, namely with higher values of $g_i$ up to $g_{1,2} \lesssim \omega_{a}$~\cite{Niemczyk10, Bayer17, Kockum19, Note_g}, to investigate the advantages we can get in the charging of the QB with respect also to the weak coupling case. Notice that we haven't considered higher values of coupling in order to avoid possible effects associated to the spectral collapse intrinsic of the considered two-photon regime~\cite{Emary02, Dolya09, Chen12, Peng17, Garbe17, Garbe19}.

\begin{figure*}[ht]
\centering
\includegraphics[width=1.8\columnwidth]{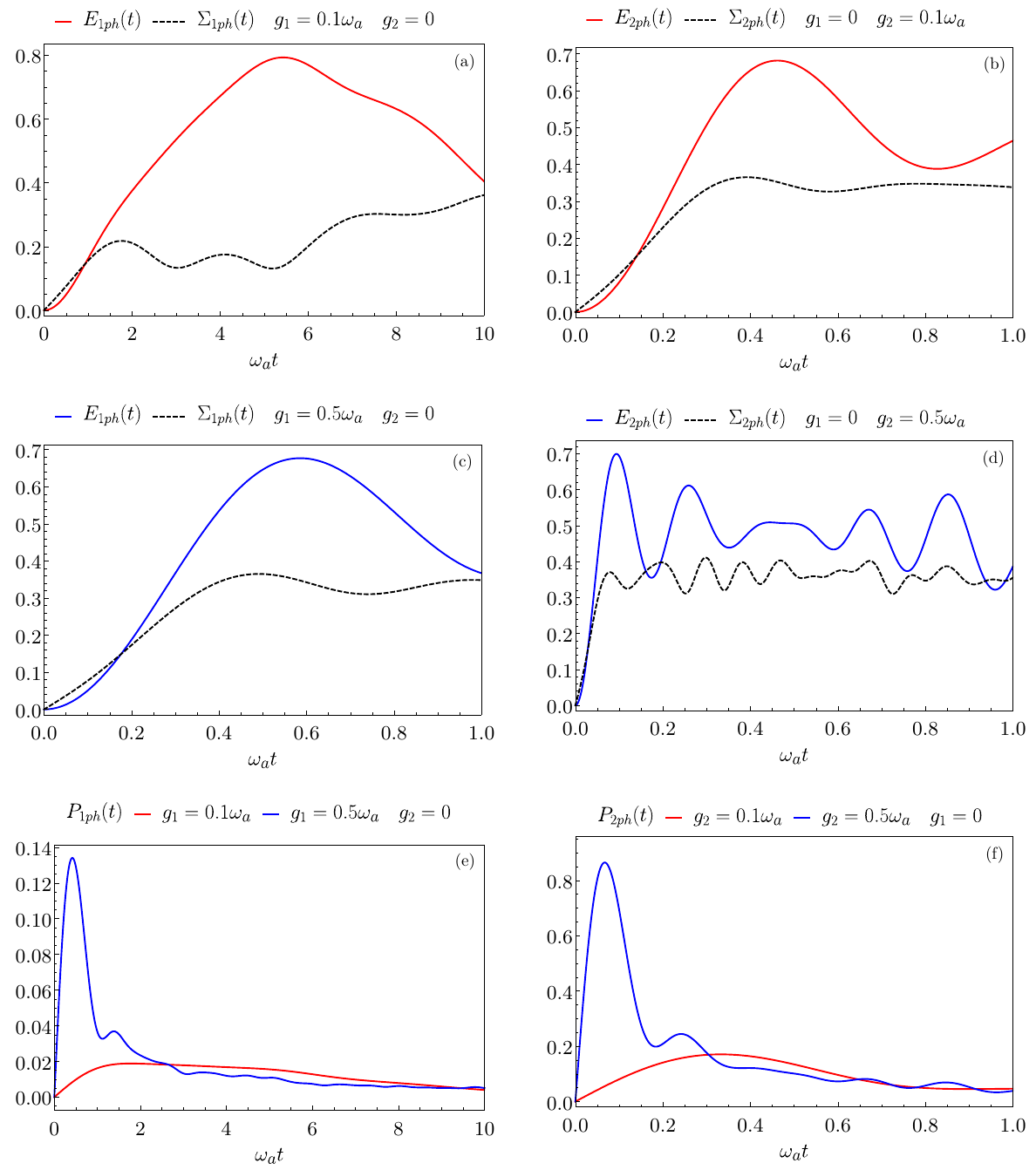}
\caption{(a)-(d) Behavior of $E(t)$ and $\Sigma(t)$ (in units of $N\omega_a$) as a function of the time  (in units of $\omega^{-1}_{a}$) for the single-photon Dicke regime with $N=10$ at (a) $g_1=0.1\omega_{a}$ and $g_2=0$, (c) $g_1=0.5\omega_{a}$ and $g_2=0$ and for the two-photon Dicke regime with $N=10$ at (b) $g_1=0$ and $g_2=0.1\omega_{a}$, (d) $g_1=0$ and $g_2=0.5\omega_{a}$. Behavior of $P(t)$ (in units of $N^2 \omega_a^2$) as a function of the time (in units of $\omega^{-1}_{a}$) for (e) the single-photon Dicke regime with $N=10$ at $g_1=0.1\omega_a$ and $g_2=0$ (red curve), $g_1=0.5\omega_{a}$ and $g_2=0$ (blue curve) and for (f) the two-photon Dicke regime with $N=10$ at $g_1=0$ and $g_2=0.1\omega_{a}$ (red curve), $g_1=0$ and $g_2=0.5\omega_{a}$ (blue curve). In panels (b) and (f) the timescale is different to properly show the maximum of the energy and charging power.}
\label{Figure4}
\end{figure*}

We start by considering the stored energy and its fluctuations in the single-photon case and by comparing them with the results obtained in the two-photon one. In Fig.~\ref{Figure4} (panels (a)-(d)) we report the behavior of $E(t)$ and $\Sigma(t)$ for the single-photon Dicke regime ($g_2=0$ in panels (a) and (c)) in comparison with the two-photon one ($g_1=0$ in panels (b) and (d)). 
The main result one can infer is that also in this case the two-photon regime ($g_1=0$) requires less time to achieve the maximal charging of the QB (see also Table \ref{tab1}). In particular, we observe that in the intermediate case of $g_{1,2}=0.1\omega_{a}$ the ratio in Eq.~(\ref{ratio_R}) becomes $\mathcal{R}\sim11.75$ while for $g_{1,2}=0.5\omega_{a}$ we obtain $\mathcal{R}\sim6.24$.
We also underline that the USC regime leads to faster charging times compared to the weak coupling case we have investigated in Section~\ref{sec:WC}, both in the single-photon and two-photon cases. Moreover we observe that within the USC regime we obtain better performances with higher coupling constant ($g_{1,2}=0.5 \omega_a$).

Concerning the energy, in general we can see that the maximum value we obtain is similar for all the considered cases.

\begin{table}[!h]
\centering
 \begin{tabular}{ | c | c | c | }
    \hline
     &   Dicke $1ph$ ($g_2=0$)  & Dicke $2ph$ ($g_1=0$)  \\ 
   \cline{2-3} &\begin{tabular}{cc}$E_{max}$&$\hspace{0.5cm} \omega_at_E$\end{tabular}& \begin{tabular}{cc} $E_{max}$ & $\hspace{0.5cm} \omega_at_E$ \end{tabular} \\ \hline
     $g_{1,2}=0.1\omega_{a}$&\begin{tabular}{cc} $\hspace{0.15cm}0.793$ & $\quad 5.431$\end{tabular} & \begin{tabular}{cc} $\hspace{0.15cm}0.681$ & $\quad 0.462$  \end{tabular}\\ \hline 
     $g_{1,2}=0.5\omega_{a}$ &\begin{tabular}{c c}$\hspace{0.15cm}0.677$ & $\quad0.580$ \end{tabular}& \begin{tabular}{c c}$\hspace{0.15cm}0.699$ & $\quad0.093$ \end{tabular} \\ \hline      
    \end{tabular}
    \caption{Maximum of the stored energy $E(t_{E})$ (in unit of $N\omega_a$) and corresponding charging time $t_{E}$ (in units of $\omega^{-1}_{a}$) for the single-photon ($g_2=0$) and two-photon ($g_1=0$) Dicke regimes for $g_{1,2}=0.1\omega_{a}$, $g_{1,2}=0.5\omega_{a}$ and $N=10$.}
    \label{tab1}
\end{table}

As we can see from Fig.~\ref{Figure4} energy fluctuations are unavoidable and finite in all the considered range of parameters and for both the considered couplings. The maximum of the correlator is around $\bar{\Sigma}/N\omega_{a}\sim 0.35$ in all the reported cases, except for the panel (a) where a better charging is achieved and consequently we obtain $\bar{\Sigma}/N\omega_{a}\sim0.14$.
This confirms the fact that interactions between $N$ TLSs mediated by the cavity radiation leads to fluctuations due to the fact that the condition $E_{max}=N\omega_a$ is never achieved.

We now consider the charging power $P(t)$ as defined in Eq.~(\ref{P(t)}) reported in Fig.~\ref{Figure4} (panels (e)-(f)).

From Table~\ref{tab2}, both the maxima of the charging power and the times at which they occur are better in the two-photon regime. In particular, the ratio in Eq. (\ref{ratio_L}) is $\mathcal{L}\sim 7.44$ for $g_{1,2}=0.1\omega_{a}$ and of $\mathcal{L}\sim 5.08$ for $g_{1,2}=0.5\omega_{a}$. This is another signature of the fact that the two-photon coupling can lead to a greater charging power with respect to the single-photon one.
Furthermore we can state that within the USC regime we also obtain an improved charging power for higher coupling constant ($g_{1,2}=0.5\omega_a$).

\begin{table}[!h]
\centering
 \begin{tabular}{ | c | c | c | }
    \hline
     &   Dicke $1ph$ ($g_2=0$)  & Dicke $2ph$ ($g_1=0$)  \\ 
   \cline{2-3} &\begin{tabular}{cc}$P_{max}$&$\hspace{0.5cm} \omega_at_P$\end{tabular}& \begin{tabular}{cc} $P_{max}$ & $\hspace{0.5cm} \omega_at_P$ \end{tabular} \\ \hline
     $g_{1,2}=0.1\omega_{a}$&\begin{tabular}{cc} $\hspace{0.15cm} 0.018$ & $\quad 1.770$\end{tabular} & \begin{tabular}{cc} $\hspace{0.15cm}0.170$ & $\quad0.328$  \end{tabular}\\ \hline 
     $g_{1,2}=0.5\omega_{a}$ &\begin{tabular}{c c}$\hspace{0.15cm} 0.134$ & $\quad 0.419$ \end{tabular}& \begin{tabular}{c c}$\hspace{0.15cm}0.865$ & $\quad0.067$ \end{tabular} \\ \hline      
    \end{tabular}
    \caption{Maximum of the charging power (in unit of $N^2\omega_a^2$) and corresponding $t_P$ (in units of $\omega^{-1}_{a}$) for the single-photon ($g_2=0$) and two-photon ($g_1=0$) Dicke regimes for $g_{1,2}=0.1\omega_{a}$, $g_{1,2}=0.5\omega_{a}$ and $N=10$.}
    \label{tab2}
\end{table} 

\begin{figure}[h!]
\centering
\includegraphics[width=0.95\columnwidth]{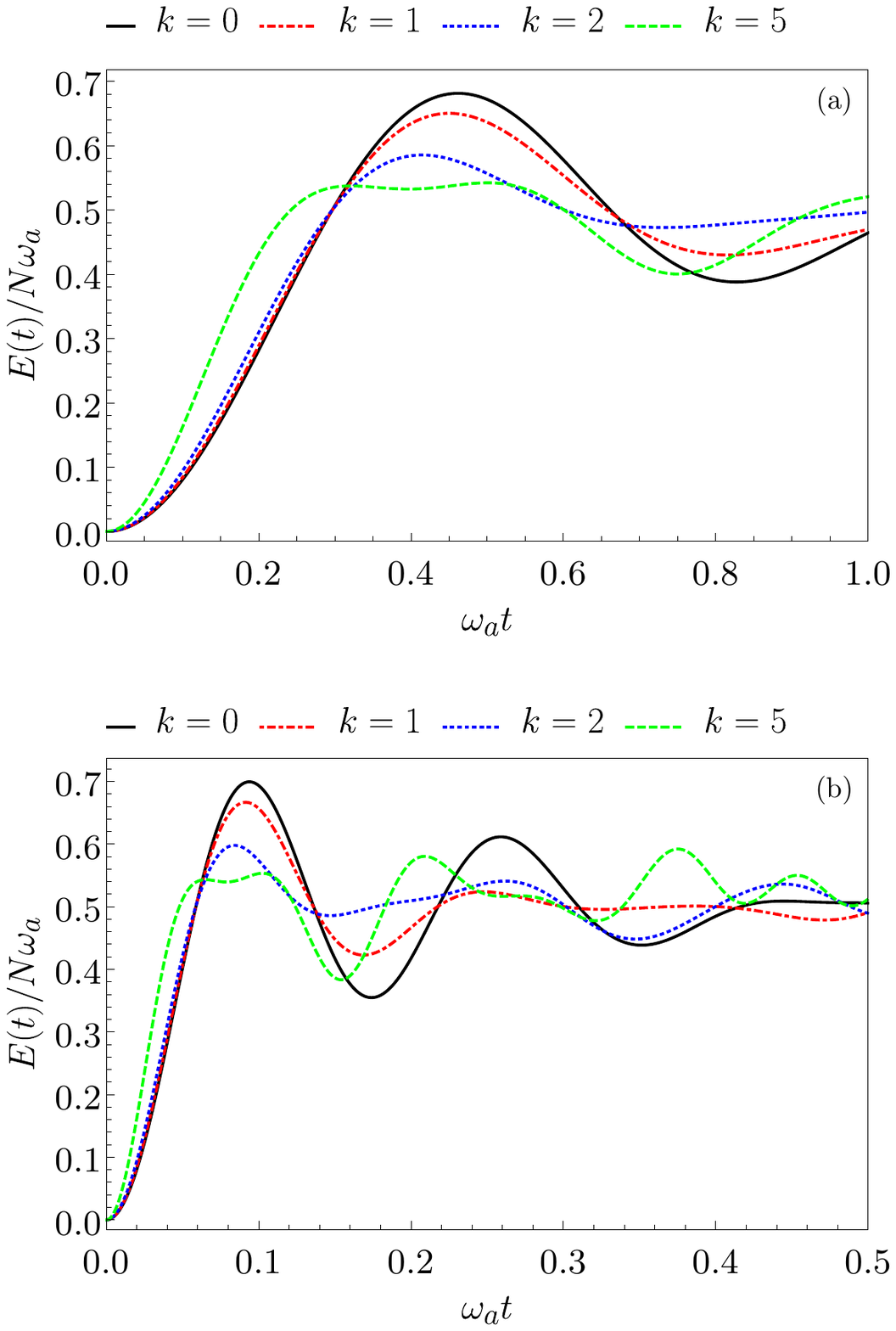}
\caption{Behavior of $E(t)$ (in units of $N\omega_a$) as a function of time (in units of $\omega^{-1}_{a}$) for the Dicke model in Eq.~(\ref{D2ph}) in correspondence of the two-photon resonance ($\omega_a=2\omega_c$) for different values of $k$, in panel (a) for $g_2=0.1\omega_{a}$ and in panel (b) for $g_2=0.5\omega_{a}$ at $N=10$.}
\label{Figure5}
\end{figure}

To conclude this section we observe that, in general, coupling constant in the USC regime lead to better performances compared to the weak coupling one both in the single-photon and two-photon coupling. 
However, to fully understand advantages and drawbacks of the USC case we analyze how the single-photon contribution can affect the results we have just obtained in the two-photon case. To do so, we consider again the total Hamiltonian in Eq.~(\ref{D2ph}) at the two-photon resonance ($\omega_a=2\omega_c$). 
In Fig.~\ref{Figure5} we report $E(t)$ for different values of $k$ (defined in Eq.~(\ref{k})) in the USC regime, where deviations from the two-photon case ($k=0$) are already observed in both panel (a) and (b) at $k=1$. 
By increasing the values of the ratio $k$ these deviations are further enhanced (see red dash-dotted curve, blue dotted curve and green dashed curve). 
Therefore, compared to the weak limit regime in Fig.~\ref{Figure3}, the presence of the single-photon interaction has a stronger impact in the USC regime. Thus in this regime, in view of actual experimental implementations, it would be necessary to properly engineer the single- and two-photon coupling to access a regime where the physics associated to the two-photon coupling clearly emerges.

\section{Collective power enhancement \label{sec:power}}

We now analyze the scaling of the maximum of the energy $E_{max}$ in Eq.~(\ref{Emax}), the maximum of the power $P_{max}$ in Eq.~(\ref{Pmax}) and the value of the energy fluctuations $\bar{\Sigma}$ at the maximum of the energy in Eq.~(\ref{barS}) as a function of the number $N$ of TLSs. We recall that in Ref.~\cite{Ferraro18} it has been shown that for the single-photon interaction the energy scales extensively with $N$, while the power shows a super-extensive behaviour with $N$, i. e. $P\propto N^{3/2}$ for large $N$. Here we focus our attention on the case of the pure two-photon coupling. We thus set $g_1=0$. Note that here the maximum of the charging power is rescaled also by the effective coupling $g_2/\omega_a$ to elucidate the results obtained at different $g_2$.

\begin{figure}[h!]
\centering
\includegraphics[width=0.87\columnwidth]{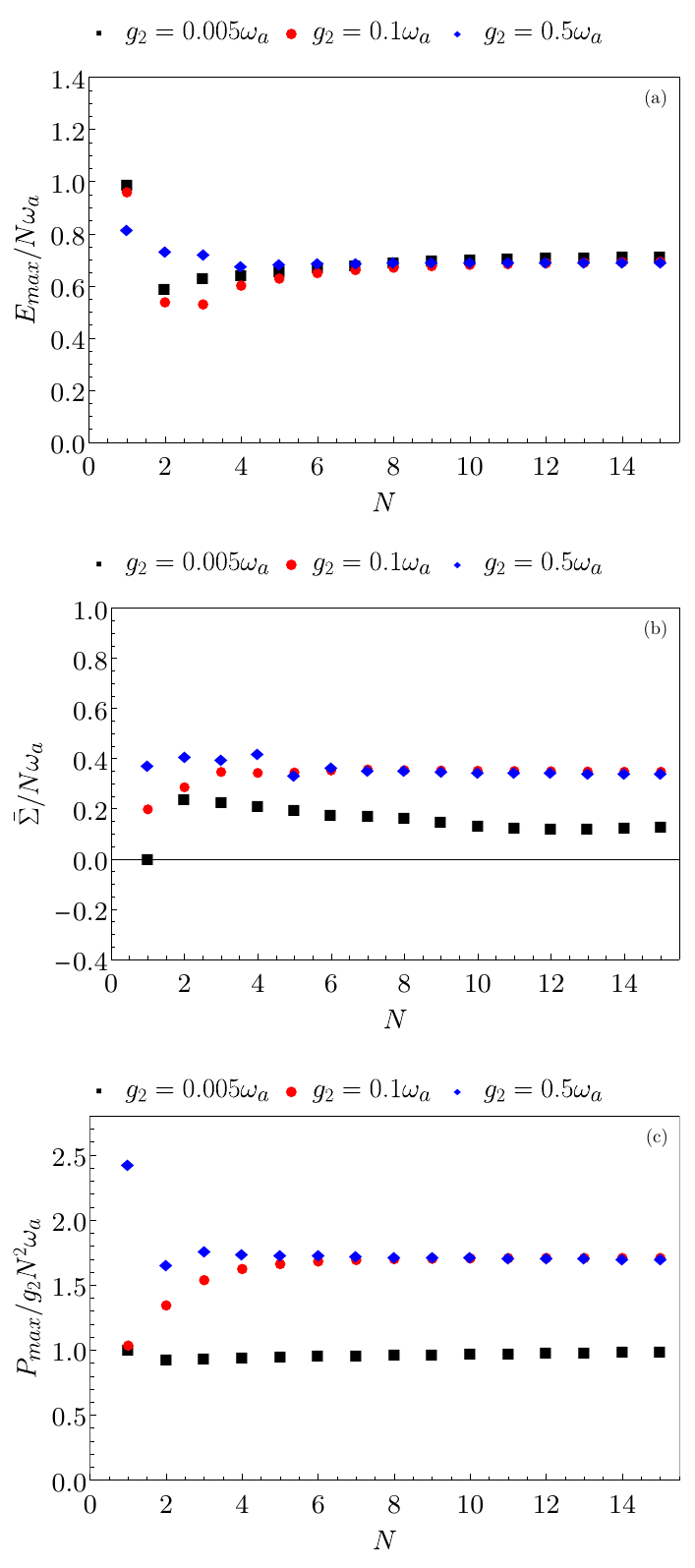}
\caption{Maximum of the energy (a) and maximum of the energy fluctuations evaluated at the maximum of the energy (b) (in units of $N \omega_a$) as a function of $N$. Maximum of the charging power (in units of $g_2 N^2\omega_a$) as a function of $N$ (c) as function of $N$. Notice that here the power is rescaled by the effective couplings $g_2/\omega_a$ to better elucidate the results at different $g_2$. Other parameters are $g_2=0.005\omega_{a}$ (black squares), $g_2=0.1\omega_{a}$ (red circles) and $g_2=0.5\omega_{a}$ (blue diamonds) and $g_1=0$ (pure two-photon coupling).}
\label{Figure6}
\end{figure}

In Fig.~\ref{Figure6} we report the above quantities as a function of $N$ for the three different values of the coupling constant considered in Section~\ref{sec:charging}. 

We note that all the quantities converge, for quite large $N$ to a steady value, reported in Table~\ref{tab3}. This consideration has been verified numerically also for higher values with respect to the ones reported in the plots (up to $N=30$).

\begin{table}[!h]
\centering
 \begin{tabular}{ | c | c | c | c | }
    \hline
     &   $E_{max}$  & $\bar{\Sigma}$  & $P_{max}$  \\ 
   \cline{1-4} 
     $g_2=0.005\omega_{a}$ & 0.730&0.140& 1.010\\ \hline 
     $g_2=0.1\omega_{a}$ & 0.700&0.350& 1.710  \\ \hline     
       $g_2=0.5\omega_{a}$ & 0.710&0.345 & 1.720\\ \hline      
    \end{tabular}
    \caption{Large $N$ steady values of $E_{max}$, $\bar{\Sigma}$ (in units of $N\omega_a$) and $P_{max}$ (in units of $g_2 N^2\omega_a$) for $g_2=0.005\omega_{a}$, $g_2=0.1\omega_{a}$ and $g_2=0.5\omega_{a}$ at $g_1=0$.}
    \label{tab3}
\end{table} 

This shows that, for large but finite value of $N$, $E_{max}$, $\bar \Sigma$ and $P_{max}$ follow the scaling laws

\begin{eqnarray}
E_{max}&\propto& N\\
\bar \Sigma&\propto& N\\
P_{max}&\propto& N^2.
\end{eqnarray}

In particular, the finite scaling obtained for the maximum charging power shows that in the two-photon case the quantum advantage related to the parallel charging in the single cavity is even greater compared to the single-photon one, where in Ref.~\cite{Ferraro18} it has been obtained $P\propto N^{3/2}$. For comparison in Appendix \ref{AppC} we demonstrate that the corresponding scaling for a parallel charging, where every TLS is coupled with a different cavity radiation, is linear in $N$ for both the maximum energy and averaged charging power. However, both in the single-photon and two-photon regimes these scalings in the thermodynamics limit don't hold and the scaling with $N$ of the charging power is recovered~\cite{Farre20, note_scaling}.

Looking at the data corresponding to $g_2=0.1\omega_{a}$ (red circles) we can note an interesting behavior. The system for low number of TLSs ($N=1,2$) behaves like it belongs to the weak coupling regime (represented by $g_2=0.005\omega_{a}$), while for larger $N$ it converges to the USC regime (represented by $g_2=0.5\omega_{a}$). This is a consequence of the fact that the system is not simply controlled by the coupling strength $g_2$ but it is greatly influenced by the number of TLSs, making the renormalized quantity $g_2 N$ the relevant control parameter for the behavior of the system as can be also deduced by the scaling of the average charging power. 

Apart from the scaling behavior $\propto N^2$, it is also interesting to remark that this trend changes with the number of TLSs for intermediate coupling strengths. Indeed, this can constitute an important hint for the development and consequent realization of QBs consisting of finite number of cells.

From Fig.~\ref{Figure6} we can also observe that, while the steady value of the energy is higher when the coupling strength is smaller, the gain in power is greater for bigger $g_2$ leading to an interesting trade-off with potential implication for practical implementations. Moreover, the system is less affected by fluctuations in the weak coupling limit.

\begin{figure}[h!]
\centering
\includegraphics[width=0.87\columnwidth]{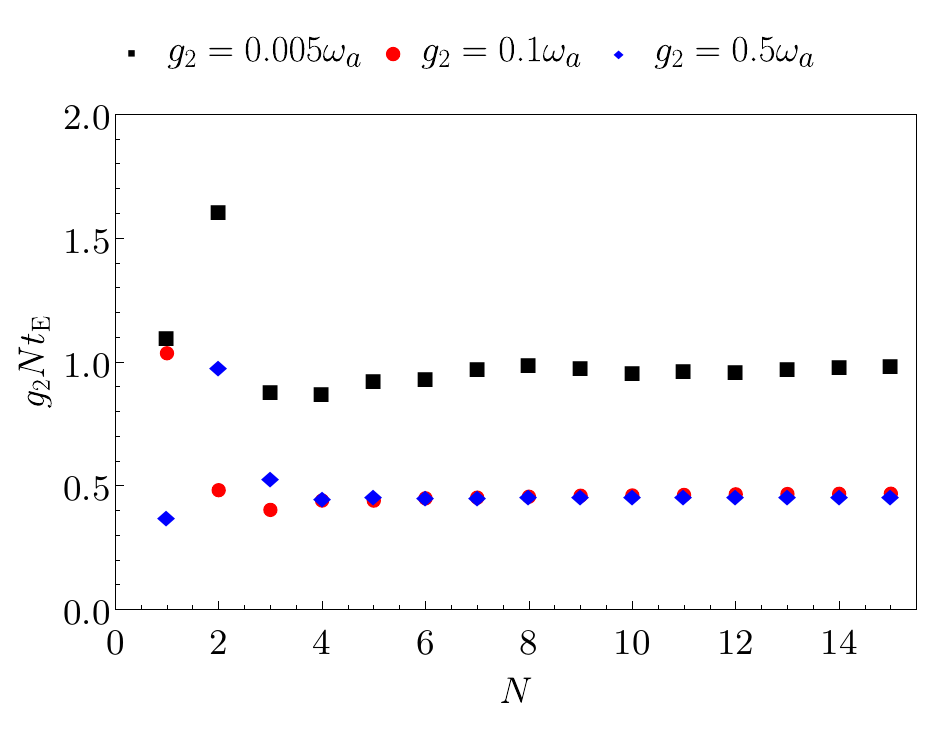}
\caption{Behaviour of $t_E$ (in units of $1/g_2N$) as a function of $N$ for the two-photon case at $g_2=0.005\omega_a$ (black squares), $g_2=0.1\omega_a$ (red circles) and for $g_2=0.5\omega_a$ (blue diamonds). }
\label{Figure7}
\end{figure}

Other important considerations about the advantage of the two-photon interaction with respect to the single-photon one can be drawn from the behavior of the charging times $t_E$ at large $N$. As we can see from Figure~\ref{Figure7}, the time obtained at the maximum of the energy scales as $t_E\propto 1/g_2N$. This means that for large $N$ the Rabi oscillations characterizing the charging time scales occurs with a characteristic time $t\propto G^{-1}$ with $G$ a renormalized many-body interaction such that $G\propto g_{2} N$ for the two-photon case. In the single-photon case \cite{Fink09} $G\propto g_{1} \sqrt{N}$, proving once more that the two-photon interaction leads to better performances of the QB compared to the single-photon one. As shown before (see Section~\ref{sec:WC}), faster charging times $t_E$ are achieved in the two-photon case even at small $N$ due to the bosonic nature of the cavity, however in this case the collective behaviour related to an increasing $N$ results in an additional speed up mechanism.

%%%%%%%%%%%%%%%%%%%%%%%%%%%%%

\section{Conclusions \label{sec:conclusions}}
We have considered a quantum battery described by a Dicke model where $N$ two-level systems interact with a cavity radiation by means of both a single- and a two-photon coupling.
We have determined the range of parameters where the former has a negligible contribution in the dynamics of the systems.
The effects of pure two-photon interaction on several figures of merits for the QB, such as the energy stored in the battery, its fluctuations and the associated charging power, has been investigated as a function of time as well as their scaling with the number of TLSs $N$. 
This kind of interaction, for a finite size system, can lead to a faster charging of the battery (scaling as $N^{-1}$) and to a consequent higher averaged charging power (scaling as $N^{2}$) with respect to the conventional single-photon contribution with relevant implication in the performance of the quantum battery. This interesting behavior can be further enhanced by moving from the weak to the ultra-strong coupling regime. 

Our analysis can be extended also to processes involving higher order photon interaction. However, such kind of non conventional coupling requires a very complex engineering to be actually implemented in realistic devices~\cite{Garziano15}.

\begin{acknowledgments}
We would like to thank G. M. Andolina and P. Scarlino for useful discussions. 
\end{acknowledgments}

%%%%%%%%%%%%%%%%%%%%%%%%%%%%%

\appendix

\section{Entries of the Hamiltonian matrix for the exact diagonalization} \label{AppA}

This Appendix is devoted to the derivation of the explicit form of the entries of the Hamiltonian matrix in Eq.~(\ref{D2ph}). Using the following relations for ladder operator of photons and pseudo-spin

\begin{eqnarray}
a^\dagger |n,l,m\rangle &=& \sqrt{n+1}  |n+1,l,m\rangle\\
a |n,l,m\rangle &=& \sqrt{n}  |n-1,l,m\rangle\\
J_\pm |n,l,m\rangle &=& \sqrt{l(l+1)-m(m\pm1)}|n,l,m\pm 1\rangle
\end{eqnarray}
together with 
\begin{eqnarray}
a^\dagger a  |n,l,m\rangle &=& n  |n,l,m\rangle \\ 
J_z  |n,l,m\rangle &=& m  |n,l,m\rangle, 
\end{eqnarray}
and recalling that $J_x=(J_++J_-)/2$, it is possible to write the matrix elements of the Dicke Hamiltonian in Eq.~(\ref{D2ph}) at positive times as

\begin{equation}\label{matrixel}\begin{array}{l}
\langle n',\frac{N}{2},\frac{N}{2}-q' |H|n,\frac{N}{2},\frac{N}{2}-q \rangle =\\\\
\frac{\omega_a}{2}\bigg(n+N-2q\bigg)\delta_{n',n}\delta_{q',q} \\\\
+g_1 \bigg[\sqrt{(n+1)[N+q(N-q-1)]}\delta_{n',n+1}\delta_{q',q+1}\\\\
+\sqrt{(n+1)[q(N-q+1)]}\delta_{n',n+1}\delta_{q',q-1}\\\\
+\sqrt{n[N+q(N-q-1)]}\delta_{n',n-1}\delta_{q',q+1}\\\\
+\sqrt{n[q(N-q+1)]}\delta_{n',n-1}\delta_{q',q-1} \bigg]\\\\
+g_2\bigg[\sqrt{(n+1)(n+2)[N+q(N-q-1)]}\delta_{n',n+2}\delta_{q',q+1}\\\\+\sqrt{(n+1)(n+2)[q(N-q+1)]}\delta_{n',n+2}\delta_{q',q-1}\\\\
+\sqrt{n(n-1)[N+q(N-q-1)]}\delta_{n',n-2}\delta_{q',q+1}\\\\
+\sqrt{n(n-1)[q(N-q+1)]}\delta_{n',n-2}\delta_{q',q-1} \bigg]
\end{array}\end{equation}
over the basis $|n,j,m\rangle$, where $n$ is the number of photons, $j(j+1)$ is the eigenvalue of $J^2$ and $m$ is the eigenvalue of $J_z$ (see main text).

As discussed in the main text, this infinite Hilbert space can be safely truncated considering a maximum number of photons $N_{ph}=4N$, higher values of photons leading only to a marginal correction of the values of the physical quantities considered.  

\section{Comparison with the Tavis-Cummings model}\label{AppB}

The TC Hamiltonian is obtained from the Dicke Hamiltonian in Eq.~(\ref{D2ph}) by performing the rotating-wave approximation and it has the following form

\begin{eqnarray} H_{TC}&=& \omega_ca^\dagger a+ \omega_a J_z + \theta(t)g_1[a^\dagger J_-+ a J_+]\nonumber \\
&+& \theta(t)g_2[(a^\dagger)^2J_-+ a^2J_+], \end{eqnarray}

\noindent where $J_\pm=J_x\pm i J_y$ and all the parameters are the same as the one in the Dicke Hamiltonian.
The advantage of this limit is that counter-rotating terms are absent, meaning that the number of excitations is always conserved and the diagonalization of the Hamiltonian is way easier than the one of the Dicke model. Indeed the matrix elements of the TC Hamiltonian can be obtained from Eq.~(\ref{matrixel}) by imposing the further constraints $n=q$ and $n' =q'$.

\begin{figure}[h!]
\centering
\includegraphics[width=0.45 \textwidth]{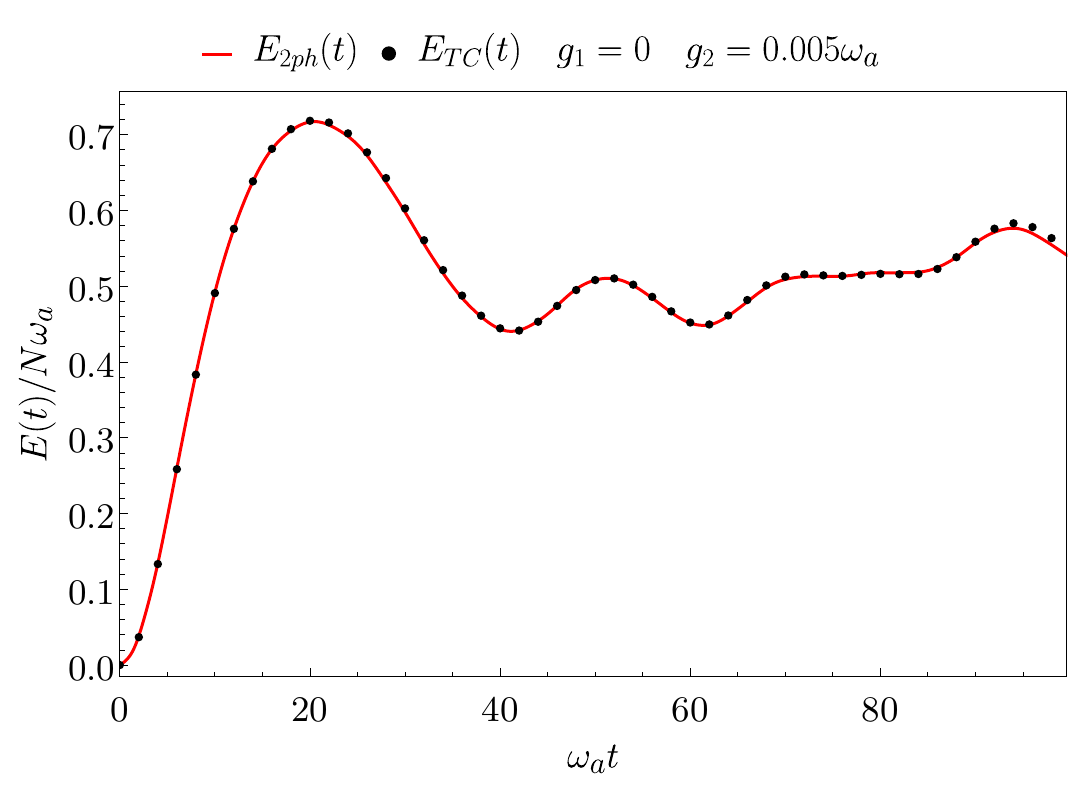}
\caption{Behaviour of $E(t)$ (in units of $N\omega_a$) as a function of time (in units of $\omega_a^{-1}$) for the two-photon Dicke model (full red curve) and for the TC model (dotted black curve) in the weak coupling regime at $g_1=0$ and $g_2=0.005\omega_a$ for $N=10$.}
\label{Figure8}
\end{figure}

This approximation is useful to further check the validity of our numerical results achieved in the limit of the weak coupling regime. Here, in Figure~\ref{Figure8}, as a representative case, we report this check for the two-photon case $g_1=0$ and $g_2=0.005\omega_a$.
As we can see the curve obtained with the numeric approach described in Appendix~\ref{AppA} and the one obtained with the TC model, just described, perfectly agree for this choice of parameters. This is also true for other values in the weak coupling regime where $g_i \ll \omega_a$ ($i=1,2$).

\section{Energy, fluctuations and average charging power for $N$ independent TLSs}\label{AppC}

We now study the stored energy, its fluctuations and the average charging power for $N$ independent TLSs in order to compare them to the collective case discussed in Section~\ref{sec:charging}. To do so, and for sake of simplicity, we show the results obtained in the framework of the two-photon Jaynes-Cummings model~\cite{JC63, Bartzis91}, valid for $N$ independent TLSs and small couplings $g_2 \ll \omega_{a}$.
We stress that the following considerations about the scaling with $N$ also hold true for greater values of the coupling. In order to make fair comparison with what discussed in the main text, here again we choose the representative value $g_2=0.005\omega_{a}$. 

We recall the Jaynes-Cummings Hamiltonian with a two-photon coupling in the resonant case ($\omega_a=2\omega_c$):

\begin{equation} 
\label{JC2} 
H_{JC}= \frac{\omega_a}{2}\bigg(a^\dagger a+\sigma_z\bigg)+ g_2[\sigma_+(a)^2+\sigma_-(a^\dagger)^2], 
\end{equation}
where we have defined $\sigma_\pm=(\sigma_x\pm i\sigma_y)/2$. This Hamiltonian can be solved analytically. Considering the basis $|\psi_{1,n}\rangle=|n,g\rangle$ and $|\psi_{2,n}\rangle=|n-2,e\rangle$ one has the matrix representation 

\begin{equation}
\label{HJC2}
H_{JC}=\begin{pmatrix}
(n-1)\frac{\omega_a}{2} && g_2\sqrt{n(n-1)}\\\\
g_2\sqrt{n(n-1)} && (n-1)\frac{\omega_a}{2}
\end{pmatrix}.
\end{equation}

Now, considering the system being initially in the ground state $|\psi_{1,n}\rangle$  with $n=2$ at the initial time $t=0$, we can write the energy at a given time $t$ for a single TLS as
\begin{equation}
\label{EJC} 
E_{JC}= \omega_a\sin^2(g_2\sqrt{n(n-1)}t). 
\end{equation}

We can also define the fluctuations of this system using Eq.~(\ref{Sigma}) as
\begin{equation}
\label{fJC2} 
\Sigma_{JC}(t)= \frac{\omega_a}{2}\bigg|\sin(2g_2\sqrt{n(n-1)}t)\bigg|. 
\end{equation}

\begin{figure}[h!]
\centering
\includegraphics[width=0.39 \textwidth]{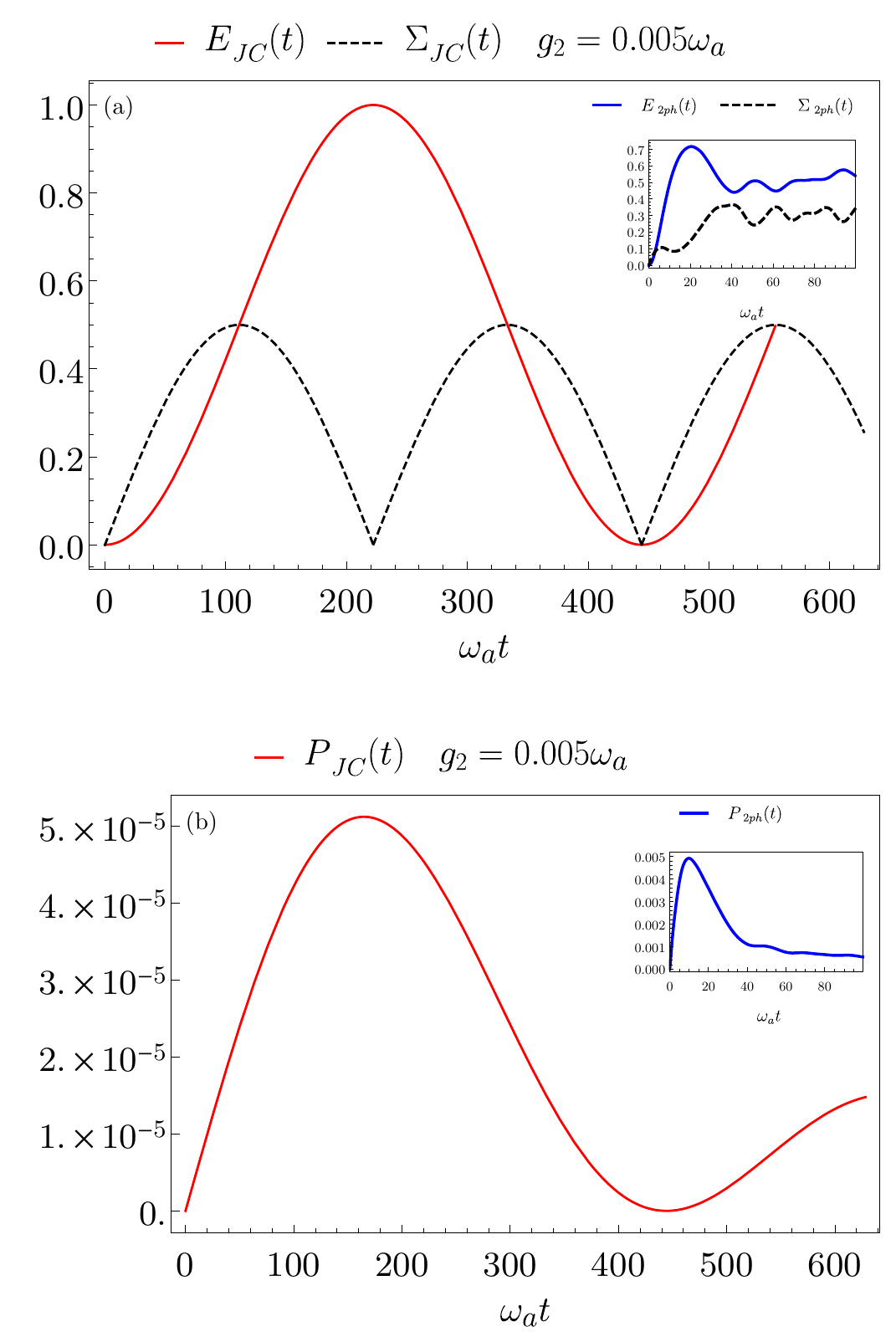}
\caption{Behavior of $E(t)$ and $\Sigma(t)$ in units of $N \omega_a$ and as a function of time (in units of $\omega^{-1}_a$ for the Jaynes-Cummings model with $n=2$ (a). Behavior of $P(t)$ in units of $N^2 \omega_a^2$ for the Jaynes-Cummings model as a function of time (in units of $\omega^{-1}_a$ with $n=2$ (b). Insets illustrates the corresponding $E(t)$, $\Sigma(t)$ (in units of $N\omega_a$) and $P(t)$ (in units of $N^2 \omega_a^2$) in the Dicke model. Other parameters are $N=10$ and $g_2=0.005\omega_{a}$.}
\label{Figure9}
\end{figure}

In Fig.~\ref{Figure9} (panel (a)) we report the energy and fluctuations for the Jaynes-Cummings model for $N=10$ independent TLSs.

\newpage

As we can see from Fig.~\ref{Figure9} and also from Eq.~(\ref{EJC}), we obtain $E_{max}=N\omega_a$. Looking at the inset in Fig.~\ref{Figure9} (panel (a)), for identical parameters the charging is worse for the Dicke model ($E_{max}^{D}\sim 0.710N\omega_a$). This shows that the interaction between the TLSs mediated by the cavity has detrimental effects on the maximal achievable stored energy.

Moreover, it is interesting to note that here the system doesn't fluctuate where the energy has its maximum, since we reach the full charging. We want to underline that with a parallel charging~\cite{Ferraro18, Andolina18} we can reach the full charging and a corresponding absence of fluctuations, however the charging times in the Dicke model are way faster (the advantage scale as $N^{-1}$) compared to the one of a QB made of independent TLSs. 
Moreover, considering the charging times, comparing the two-photon and single-photon Jaynes-Cummings model we obtain that $t_E^{2ph}/t_E^{1ph}=\sqrt{n-1}$. This justifies the fact that also in the case of one TLS the two-photon interaction can lead to better charging performance.

In addition, the major disadvantage of the JC QB is the amount of power that we can obtain from it. 
If one considers the charging power of $N=10$ independent TLSs reported in Fig.~\ref{Figure9} (b) the value of the maximum obtained in the JC model is only $P_{max}^{(JC)}\sim 5 \cdot 10^{-5}$. 
This shows how a collective charging, reported in the inset of Fig.~\ref{Figure9} (b), can enhance by a great factor ($\sim 100$) the charging power of the QB. Moreover, the maximum average charging power in the JC limit is obtained for longer times compared to the collective case.

\newpage

%%%%%%%%%%%%%%%%%%%%%%%%%%%%%


\begin{thebibliography}{100}
\bibitem{Note_power} The reason of this rescaling of the power with respect to $N^2$ will be clearer in the following.
\bibitem{Note_g} We use the short notation $g_{1,2}$ to indicate both a finite $g_{1}$ in correspondence to $g_{2}=0$ and a finite $g_{2}$ for $g_{1}=0$.
\bibitem{note_scaling} To obtain the scaling of the charging power with $N$ in the thermodynamic limit, it is necessary to renormalize the coupling constants in the Hamiltonian~(\ref{D2ph}). In particular, the single-photon coupling constant gets renormalized as $g_1/\sqrt{N}$ while the two-photon one becomes $g_2/N$. These different rescalings are related to the interaction terms of the Hamiltonian. In fact the single-photon process is connected to the electric-dipole interaction. Therefore, the electric field quantized inside a cavity is proportional to $1/ \sqrt{L}$ with $L$ the length of the cavity. In order to have a consistent thermodynamic limit, also the number of qubits have to grow with $L$ ($L \propto N$) leading to the $1/\sqrt{N}$ renormalization of $g_1$. In contrast, the two-photon interaction term of the Hamiltonian is quadratic in the electric field, with a consequent renormalization $1/N$ of the coupling strength $g_2$.
\end{thebibliography}
\end{document}